\documentclass[aps,twocolumn,prb,floatfix]{revtex4}
\usepackage{graphicx}
\usepackage{latexsym}
\usepackage{amsmath}
\usepackage{graphics}
\usepackage{amssymb}
\usepackage{layout}
\usepackage{verbatim}
\usepackage{amsfonts,epsfig}
\newcommand{\ket}[1]{\left|#1\right>}
\newcommand{\bra}[1]{\left< #1 \right|}
\begin{document}
\title{Effect of quantum fluctuations on even-odd energy difference in a Cooper-pair box.}
\author{R. M. Lutchyn}
\affiliation{ W.I.\ Fine Theoretical Physics Institute, University
of Minnesota, Minneapolis, Minnesota 55455, USA}
\date{\today }

\begin{abstract}
We study the effect of quantum charge fluctuations on the discrete
spectrum of charge states of a  small superconducting island
(Cooper-pair box) connected to a large finite-size superconductor
by a tunnel junction. In particular, we calculate the reduction of
the even-odd energy difference $\delta E$ due to virtual tunneling
of electrons across the junction. We show that the renormalization
effects are important for understanding the quasiparticle
``poisoning" effect because $\delta E$ determines the activation
energy of a trapped quasiparticle in the Cooper-pair box. We find
that renormalization of the activation energy depends on the
dimensionless normal-state conductance of the junction $g_{_T}$,
and becomes strong at $g_{_T}\gg 1$.
\end{abstract}

\pacs{03.67.Lx, 03.65.Yz, 74.50.+r, 85.25.Cp}

\maketitle

%\section{Introduction.}

Recently, superconducting quantum circuits have attracted
considerable interest (see~[\onlinecite{Wendin, Makhlin}] and
references therein). From the viewpoint of quantum many-body
phenomena, these circuits are good systems to study the effect of
quantum fluctuations of an environment on the discrete spectrum of
charge states~\cite{Lehnert, Joyez, Rimberg, Duty, Sillanpaa}
(similar to the Lamb shift in a hydrogen atom). While most of the
studies of superconducting nanostructures focus on smearing of the
charge steps in the Coulomb staircase measurements~\cite{Glazman},
here we consider another observable quantity - even-odd-electron
energy difference $\delta E$ in the Cooper-pair box (CPB). This
quantity is important for understanding the quasiparticle
``poisoning" effect~\cite{Mannik, Aumentado, Turek, Guillaume,
Matveev}, and it has been recently studied
experimentally~\cite{Naaman, Ferguson}. It was conjectured that
$\delta E$ may be reduced in the strong tunneling regime
$g_{_T}\!=\!R_q/R_{N}> 1$ by quantum fluctuations of the
charge~\cite{Naaman}. Here $R_q$ and $R_{N}$ are the resistance
quantum, $R_q=h/e^2$, and normal-state resistance of the tunnel
junction, respectively.

In this paper, we study the renormalization of the discrete
spectrum of charge states of the Cooper-pair box by quantum charge
fluctuations. We show that virtual tunneling of electrons across
the tunnel junction may lead to a substantial reduction of the
even-odd energy difference $\delta E$. We consider here the case
of the tunnel junction with a large number of low transparency
channels~\cite{applicability}.

\begin{figure}
\centering
\includegraphics[width=2.8in]{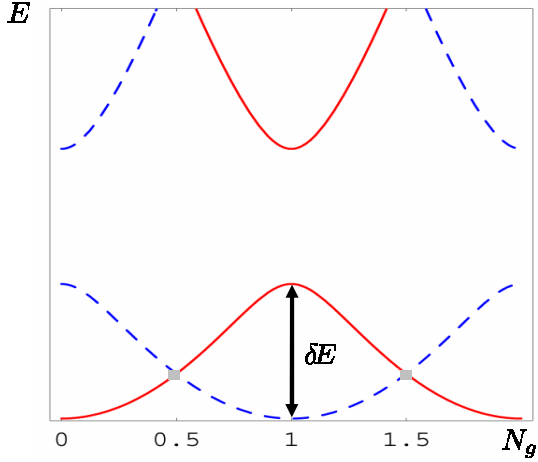}
\caption{(color online). Energy of the Cooper-pair box as a
  function of dimensionless gate voltage $N_g$ in units of $e$.
  Solid (red) line corresponds to even-charge state of the box, dashed (blue) line
  corresponds to the odd-charge state of the box. Here $\delta
  E$ is the ground state energy difference between the even-charge state (no
quasiparticles in the CPB), and odd-charge state (an unpaired
electron in the CPB) at $N_g=1$. (We assume here equal gap
energies in the box and reservoir,
$\Delta_{r}=\Delta_{b}=\Delta$.)} \label{fig1}
 \end{figure}

The dynamics of the system is described by the Hamiltonian
\begin{equation}\label{Hqubit}
H=H_{_C}+H^b_{\rm{BCS}}+H^r_{\rm{BCS}}+H_{_T}.
\end{equation}
Here $H^b_{\rm{BCS}}$ and $H^r_{\rm{BCS}}$ are BCS Hamiltonians
for the CPB and superconducting reservoir;
$H_{_C}=E_c(\hat{Q}/e-N_g)^2$ with $E_c$, $N_g$  and $\hat{Q}$
being the charging energy, dimensionless gate voltage and charge
of the CPB, respectively. The tunneling Hamiltonian $H_{_T}$ is
defined in the conventional way. We assume that the island and
reservoir are isolated from the rest of the circuit; \emph{i.e.}
total number of electrons in the system is fixed. At low
temperature $T<T^*$, thermal quasiparticles are frozen out. (Here
$T^*=\frac{\Delta}{\ln(\Delta/\delta)}$ with $\Delta$ and $\delta$
being superconducting gap and mean level spacing in the reservoir,
respectively). If total number of electrons in the system is even,
then the only relevant degree of freedom at low energies is the
phase difference across the junction $\varphi$. In the case of an
odd number of electrons a quasiparticle resides in the system even
at zero temperature. The presence of $1e$-charged carriers changes
the periodicity of the CPB energy spectrum (see Fig.~\ref{fig1})
since an unpaired electron can reside in the island or in the
reservoir. Note that at $N_g=1$, a working point for the charge
qubit, the odd-electron state of the CPB may be more favorable
resulting in trapping of a quasiparticle in the
island~\cite{Naaman, Ferguson, lutchyn3}. In order to understand
energetics of this trapping phenomenon, one has to look at the
ground state energy difference $\delta E$ between the even-charge
state (no quasiparticles in the CPB), and odd-charge state (with a
quasiparticle in the CPB):
\begin{eqnarray}\label{activation}
\delta E=E_{\rm{even}}(N_g\!=\!1)-E_{\rm{odd}}(N_g\!=\!1),
\end{eqnarray}
see also Fig.~\ref{fig1}. For equal gap energies in the box and
the reservoir ($\Delta_{r}=\Delta_{b}=\Delta$) the activation
energy $\delta E$ is determined by the effective charging energy
of the CPB. Note that tunneling of an unpaired electron into the
island shifts the net charge of the island by $1e$. Thus, one can
find $\delta E$ of Eq.~(\ref{activation}) as the energy difference
at two values of the induced charge, $N_g=1$ and $N_g=0$, on the
even-electron branch of the spectrum (see Fig.~\ref{fig1}):
\begin{eqnarray}\label{evenodd}
\delta
E\!&\!=\!&\!E_{\rm{even}}(N_g\!=\!1)\!-\!E_{\rm{even}}(N_g\!=\!0).
\end{eqnarray}
Here we assumed that subgap conductance due to the presence of an
unpaired electron is negligible~\cite{Lutchyn2}.

In order to find activation energy $\delta E$ given by
Eq.~(\ref{evenodd}), we calculate the partition function $Z(N_g)$
for the system, island and reservoir, with even number of
electrons. For the present discussion it is convenient to
calculate the partition function using the path integral
description developed by Ambegaokar, Eckern and
Sch\"{o}n~\cite{AES}. In this formalism the quadratic in $\hat{Q}$
interaction in Eq.~(\ref{Hqubit}) is decoupled with the help of
Hubbard-Stratonovich transformaion by introducing an auxiliary
field $\varphi$ (conjugate to the excess number of Cooper pairs on
the island). Then, the fermion degrees of freedom are traced out,
and around the BCS saddle-point the partition function becomes
\begin{eqnarray}\label{part}
Z(N_g)\!=\!\sum_{m=\!-\!\infty}^{\infty} e^{i \pi N_g
m}\int\!d\varphi_0\!\int_{\varphi(0)\!=\!\varphi_0}^{\varphi(\beta)\!=\!\varphi_0\!+\!2\pi
m}D\varphi(\tau)e^{-S}.\nonumber\\
\end{eqnarray}
Here the summation over winding numbers accounts for the
discreteness of the charge~\cite{Schon}, and the action $S$ reads
($\hbar=1$)
\begin{eqnarray}\label{AES}
S&=&\int_0^{\beta}d\tau\left[\frac{C_{\rm{geom}}}{2}\left(\frac{\dot{\varphi}(\tau)}{2e}\right)^2\!-\!E_{_J}\cos\varphi(\tau)\right]\\
&+&\int_0^{\beta}d\tau\int_0^{\beta}d\tau'\alpha(\tau\!-\!\tau')\left(1\!-\!\cos\left(\frac{\varphi(\tau)\!-\!\varphi(\tau')}{2}\right)\right)\nonumber
\end{eqnarray}
with $\beta$ being the inverse temperature, $\beta=1/T$. Here
 $C_{\rm{geom}}$ is the geometric capacitance of
the CPB which determines the bare charging energy
$E_c=e^2/2C_{\rm{geom}}$; and $E_{_J}$ is Josephson coupling given
by Ambegaokar-Baratoff relation. The last term in Eq.~(\ref{AES})
accounts for single electron tunneling with kernel $\alpha(\tau)$
decaying exponentially at $\tau\gg\Delta^{-1}$
~[\onlinecite{AES}]. For sufficiently large capacitance the
evolution of the phase is slow in comparison with $\Delta^{-1}$,
and we can simplify the last term in Eq.~(\ref{AES})
\begin{eqnarray}\label{approx1}
\int_0^{\beta}\!d\tau\!\int_0^{\beta}\!d\tau'\!\alpha(\tau\!-\!\tau')\!\left(1\!-\!\cos\left(\frac{\varphi(\tau)\!-\!\varphi(\tau')}{2}\right)\right)\approx\nonumber\\
\approx\frac{3\pi^2}{128}\frac{1}{2\pi e^2 R_{N} \Delta
}\int_0^{\beta}d\tau \left(\frac{d\varphi(\tau)}{d\tau}\right)^2.
\end{eqnarray}
It follows from here that virtual tunneling of electrons between
the island and reservoir leads to the renormalization of the
capacitance~\cite{AES}
\begin{eqnarray}\label{capacitance}
C_{\rm{geom}}\rightarrow\tilde{C}=C_{\rm{geom}}+\frac{3\pi}{32}\frac{1}{R_{N}\Delta}.
\end{eqnarray}
Within the approximation~(\ref{approx1}), the effective action
acquires simple form
\begin{eqnarray}\label{effaction}
S_{\rm{eff}}=\int_0^{\beta}d\tau\left[\frac{\tilde{C}}{2}\left(\frac{\dot{\varphi}(\tau)}{2e}\right)^2-E_{_J}\cos\varphi(\tau)\right].
\end{eqnarray}
To calculate $Z(N_g)$ one can use the analogy between the present
problem and that of a quantum particle moving in a periodic
potential, and write the functional integral as a quantum
mechanical propagator from $\varphi_i=\varphi_0$ to
$\varphi_f=\varphi_0+2\pi m$ during the (imaginary) ``time"
$\beta$
\begin{eqnarray}\label{propagator}
\int_{\varphi(0)\!=\!\varphi_0}^{\varphi(\beta)\!=\!\varphi_0\!+\!2\pi
m}\!\!D\varphi(\tau)\exp(-S_{\rm{eff}})\!=\!\bra{\varphi_0}\!e^{-\beta
\hat{H}_{\rm{eff}}}\ket{\varphi_0\!+\!2\pi m}.\nonumber\\
\end{eqnarray}
The time-independent ``Shr\"{o}dinger equation" corresponding to
such problem has the form~\cite{Kamenev}
\begin{eqnarray}\label{Matt}
\hat{H}_{\rm{eff}}\Psi(\varphi)\!=\!E\Psi(\varphi),\,
\hat{H}_{\rm{eff}}\!=\!\left(
-4\tilde{E}_c\frac{\partial^2}{\partial \varphi^2} \!-\!
E_{_J}\cos\varphi\right).\nonumber\\
\end{eqnarray}
Here $\tilde{E}_c$ denotes renormalized charging energy
\begin{eqnarray}\label{Ec}
\tilde{E}_c=\frac{E_c}{1+\frac{3}{32}g_{_T}\frac{E_c}{\Delta}}.
\end{eqnarray}
One can notice that Eq.~(\ref{Matt}) corresponds to well-known
Mathieu equation, for which eigenfunctions $\Psi_{k,s}(\varphi)$
are known~\cite{Likharev}. Here quantum number $s$ labels Bloch
band ($s=0,1,2,...$), and $k$ corresponds to the
``quasi-momentum". By rewriting the propagator~(\ref{propagator})
in terms of the eigenfunctions of the Shr\"{o}dinger
equation~(\ref{Matt}) we obtain
\begin{eqnarray}\label{9}
\bra{\varphi_0}\!&\!e^{-\beta
\hat{H}_{\rm{eff}}}\!&\!\ket{\varphi_0\!+\!2\pi m}\\\nonumber\\
\!&\!=\!&\!\sum_{k,k'}\bra{\varphi_0}\!\!\!\ket{\,k}\!\bra{k}e^{-\beta
\hat{H}_{\rm{eff}}}\ket{k'}\!\bra{k'}\!\!\!\ket{\varphi_0\!+\!2\pi m}\nonumber\\
\!&\!=\!&\!\sum_{k,s}\Psi^{*}_{k,s}(\varphi_0)\Psi_{k,s}(\varphi_0\!+\!2\pi
m)\exp\!\left(-\beta E_{s}(k)\right)\!.\nonumber
\end{eqnarray}
Here $E_{s}(k)$ are eigenvalues of Eq.~(\ref{Matt}).

\begin{figure}
\centering
\includegraphics[width=3.2in]{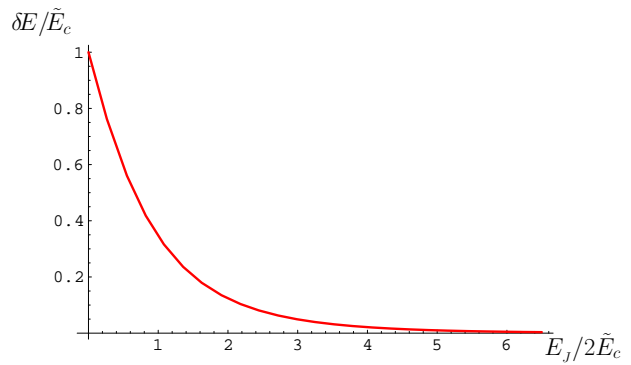}
\caption{Dependence of the even-odd energy difference $\delta E$
on the dimensionless parameter $E_{_J}/2\tilde{E}_c$.}
\label{fig2}
\end{figure}

According to the Bloch theorem, the eigenfunctions should have the
form $\Psi_{k,s}(\varphi)=e^{ik\varphi/2}u_{k,s}(\varphi)$ with
$u_{k,s}(\varphi)$ being $2\pi$-periodic functions,
$u_{k,s}(\varphi)=u_{k,s}(\varphi+~2\pi)$. We can now rewrite
Eq.~(\ref{part}) as
\begin{eqnarray}\label{answer}
Z(N_g)\!&=&\!\sum_{m=-\infty}^{\infty}\!e^{i \pi N_g m
}\!\int d\varphi_0\times\nonumber\\
&\times&\sum_{k,s}\Psi^{*}_{k,s}(\varphi_0)\Psi_{k,s}(\varphi_0\!+\!2\pi
m)\exp\left(-\beta E_{s}(k)\right)\nonumber\\
&=&\sum_{s=0,1}^{\infty}\exp\left(-\beta
E_{s}\left(N_g\right)\right).
\end{eqnarray}
The eigenvalues $E_{s}(N_g)$ are given by the Mathieu
characteristic functions $M_A(r,q)$ and $M_B(r,q)$~\cite{Cottet}.
At $N_g=0$ and $N_g=1$, the exact solution for the lowest band
reads
\begin{eqnarray}\label{band}
E_{0}(N_g=0)\!&=&\tilde{E}_c
M_{A}\!\left(0,\!-\frac{E_{_J}}{2\tilde{E}_c}\right),\nonumber\\\\
E_{0}(N_g=1)\!&=&\tilde{E}_c
M_{A}\!\left(1,\!-\frac{E_{_J}}{2\tilde{E}_c}\right).\nonumber
\end{eqnarray}
The activation energy $\delta E$ can be calculated from
Eq.~(\ref{answer}) by evaluating free energy at $T=0$:
\begin{eqnarray}\label{evenodd2}
\delta E= \tilde{E}_c\left[
M_{A}\!\left(1,\!-\frac{E_{_J}}{2\tilde{E}_c}\right)\!-\!M_{A}\!\left(0,\!-\frac{E_{_J}}{2\tilde{E}_c}\right)\right].
\end{eqnarray}
The plot of $\delta E$ as a function of $E_{_J}/2\tilde{E}_c$ is
shown in Fig.~(\ref{fig2}). Even-odd energy difference $\delta E$
has the following asymptotes:
\begin{eqnarray}\label{evenodd1}
\delta E\!\approx\!\left\{
\begin{array}{rcl}
\!\tilde{E}_c\!-\frac{1}{2}E_{_J},
\,\,\,\,\,\,\,\,\,\,\,\,\,\,\,\,\,\,\,\,\,\,\,\,\,\,\,\,\,\,\,\,\,\,\,\,\,\,\,\,\,\,\,\,\,\,\,\,\,\!\!&\!\!E_{_J}/2\tilde{E}_c\!\!&
\!\ll\!1
, \\\nonumber\\\nonumber\\
2^5\sqrt{\frac{2}{\pi}}\tilde{E}_c\left(\frac{E_{_J}}{2\tilde{E}_c}\right)^{3/4}\exp\!\left(-4\sqrt{\frac{E_{_J}}{2\tilde{E}_c}}\,\right)\!,\,
\!&\!E_{_J}/2\tilde{E}_c\!\!&
\!\gg\!1.\\
\end{array}
\right.
\end{eqnarray}
These asymptotes can be also obtained using perturbation theory
and WKB approximation, respectively.

As one can see from Eq.~(\ref{evenodd2}),  $\delta E$ can be
reduced by quantum charge fluctuations. For realistic experimental
parameters~\cite{Naaman} $\Delta\approx 2.5$K, $E_c\approx 2$K and
$g_{_T}\!\approx\!2$, we find that even-odd energy difference
$\delta E$ is $15\%$ smaller with respect to its bare value,\emph{
i.e.} $\delta E\approx 1.45$K and $\delta E^{\rm{bare}}\approx
1.7$K. Since the reduction of the activation energy by quantum
fluctuations is much larger than the temperature, this effect can
be observed experimentally. The renormalization of $\delta E$ can
be studied systematically by decreasing the gap energy $\Delta$,
which can be achieved by applying magnetic field
$B$~[\onlinecite{Lehnert}]. The dependance of the activation
energy $\delta E$ on $\Delta(B)$ in Eq.~(\ref{evenodd2}) enters
through the Josephson energy $E_{_J}$, which is given by
Ambegaokar-Baratoff relation, and renormalized charging energy
$\tilde{E}_c$ of Eq.~(\ref{Ec}).

The renormalization of the discrete spectrum of charge states in
the CPB becomes more pronounced in the strong tunneling regime.
However, the adiabatic approximation leading to the effective
action $S_{\rm{eff}}$~(\ref{effaction}) is valid when the
evolution of the phase is slow, \emph{i.e.} the adiabatic
parameter $\omega_{_J}/\Delta$ is small. (Here $\omega_{_J}$ is
the plasma frequency of the Josephson junction,
$\omega_{_J}\sim\sqrt{E_{c}E_{_J}}$.) Thus, at large conductances
$g_T$ the adiabatic approximation holds only when the geometric
capacitance is large $C_{\rm{geom}}\gg e^2g_{_T}/\Delta$. Under
such conditions the renormalization effects lead to a small
correction of the capacitance, see Eq.~(\ref{capacitance}). If
$\omega_{_J}/\Delta> 1$, the dynamics of the phase is described by
the integral equation~(\ref{AES}), and retardation effects have to
be included.

In the similar circuit corresponding to the Cooper-pair box
qubit~\cite{Wendin, Makhlin} it is possible to achieve strong
tunneling regime $g_{_T}\gg \Delta C_{\rm{geom}}/e^2$, and satisfy
the requirements for adiabatic approximation ($\omega_{_J}/\Delta
\ll 1$). In this circuit a single Josephson junction is replaced
by two junctions in a loop configuration~\cite{Wendin,Makhlin}.
This allows to control the effective Josephson energy using an
external flux $\Phi_x$. (For the CPB qubit the Josephson energy
$E_{_J}$ in Eq.~(\ref{effaction}) should be replaced with
$E_{_J}(\Phi_x)=2E^0_{_J}\cos\left(\pi\Phi_x/\Phi_0\right)$; here
$\Phi_0$ is the magnetic flux quantum, $\Phi_0=h/2e$, and
$E^0_{_J}$ is the Josephson coupling per junction.) In this setup
even at large conductance $g_{_T}\gg \Delta C_{\rm{geom}}/e^2$ one
can decrease $\omega_{J}\sim \sqrt{E_c E_{_J}(\Phi_x)}$ by
adjusting the external magnetic flux to satisfy
$\omega_{J}/\Delta\ll 1$. Under such conditions the quantum
contribution to the capacitance $\tilde{C}$ (see
Eq.~(\ref{capacitance})) becomes larger than the geometric one,
while the dynamics of the phase is described by the simple action
of Eq.~(\ref{effaction}). It would be interesting to study
experimentally the renormalization of the discrete energy spectrum
of the qubit in this regime. We propose to measure, for example,
the even-odd energy difference $\delta E$. In this case $\delta E$
is determined by the conductance of the junctions $g_{_T}$,
superconducting gap $\Delta$, and magnetic flux $\Phi_x$, and is
given by Eq.~(\ref{evenodd2}) with $\tilde{E}_c\approx
32\Delta/3g_{_T}$, see Eq.~(\ref{Ec}), and
$E_{_J}=2E^0_{_J}\cos\left(\pi\Phi_x/\Phi_0\right)$.

In conclusion, we studied the renormalization of the discrete
spectrum of charge states of the Cooper-pair box by virtual
tunneling of electrons across the junction. In particular, we
calculated the reduction of even-odd energy difference $\delta E$
by quantum charge fluctuations. We showed that under certain
conditions the contribution of quantum charge fluctuations to the
capacitance of the Cooper-pair box may become larger than the
geometric one. We propose to study this effect experimentally
using the Cooper-pair box qubit.

\begin{acknowledgments}

The author is grateful to L. Glazman and A.~Kamenev for
stimulating discussions. This work is supported by NSF grants DMR
02-37296, and DMR 04-39026.
\end{acknowledgments}

\end{document}